\documentclass[12pt,preprint]{aastex}

\def\etal{et al.\ }

\shorttitle{Starburst-ISM Interaction in NGC 1569 II.}
\shortauthors{Buckalew et al.}

\received{2005 December 13}
\begin{document}

\title{THE STARBURST-INTERSTELLAR MEDIUM INTERACTION IN NGC 1569 
II. SMALL-SCALE EXAMINATION OF NEBULAR EMISSION, H II REGION SIZE DISTRIBUTION, AND H II REGION LUMINOSITY FUNCTION}

\author{B{\sc rent} A.\ B{\sc uckalew}\altaffilmark{1,2} \& H{\sc enry} A.\ K{\sc obulnicky}\altaffilmark{1}}
\altaffiltext{1}{Department of Physics \& Astronomy, University of Wyoming, Laramie, WY  82071, {\tt chipk@uwyo.edu}}
\altaffiltext{2}{Present Address: Infrared Processing and Analysis Center, California Institute of Technology, Pasadena, CA  91125, {\tt brentb@ipac.caltech.edu}}

\begin{abstract}
As the nearest dramatic example of a poststarburst galaxy driving a
galactic wind, NGC~1569 is an ideal test environment to understand the
impact of ``feedback'' from massive star lives and deaths on the
surrounding ISM.  
We present {\it HST WFPC2}\footnotemark~narrowband imagery of NGC 1569 in an attempt to understand the underlying 
ionizing emission mechanisms on a 3 pc scale and to generate a H~{\sc ii} 
region size distribution and luminosity function.  
We use [O~{\sc iii}]/H$\beta$ and [S~{\sc ii}]/H$\alpha$ ratio maps to find 
that non-photoionizing mechanisms (e.g.\ shocks) are responsible for 10$\pm$3\% 
 of the H$\alpha$ emission, 2.3--3.3 times larger than results from similar 
galaxies.  Note that our method of determining this result is different than
these past results, a point that we discuss further in the article.  
The area of the non-photoionized region is 10--23\% of the total.  
Our results for NGC 1569 indicate that these non-photoionized areas do not lie 
in low surface brightness regions exclusively.  
A comparison with multiwavelength point source catalogs of NGC 1569 indicates 
that the dominant non-photoionizing mechanisms are shocks from 
supernovae or winds from massive stars.  
To explain this large percentage of non-photoionized emission, we suggest that 
NGC 1569 is, indeed, in a post-starburst phase as previous authors have claimed.
We also derived slopes for the H~{\sc ii} region luminosity function (-1.00$\pm$0.08) 
and size distribution (-3.02$\pm$0.27). 
The luminosity slope, though shallow, is similar to previous work on this 
galaxy and other irregular galaxies.  The size distribution slope is shallower than previous slopes
found for irregular galaxies, but our slope value fits into their confidence intervals and vice versa.  
Within 4 pc of the 10-20 Myr old superstar clusters A1, A2, 
and B, no bright
HII regions exist to a luminosity limit of 2.95$\times10^{36}$ erg s$^{-1}$,
suggesting that the winds and shocks have effectively terminated star
formation in this small cavity. 
In the three annular regions around 
the SSCs, both the HII region luminosity function and HII region size
distribution are consistent with respect to one another and the galaxy as a whole.  
The HII region surface densities within the annuli remain the same as the annuli are moved
away from the superstar clusters.  
These results indicate that ``feedback'' effects in NGC~1569 are confined to the
immediate vicinity of the most recent massive star formation event on
scales of $\sim1$ pc.  
\end{abstract}

\keywords{galaxies:  individual (NGC~1569) --- galaxies: ISM --- galaxies: structure --- ISM: structure}

\footnotetext{Based on observations with the NASA/ESA {\it Hubble Space Telescope}, 
obtained at the Space Telescope Science Institute, which is operated by AURA, Inc., 
under NASA contract NAS5-26555.}

\section{INTRODUCTION}

\subsection{Feedback Mechanisms} 
Elevated star formation rates of starburst galaxies produce a corresponding increase in the numbers of massive stars and supernovae.  Stellar winds and supernovae can ionize their surrounding interstellar medium (ISM) to levels that cannot be explained by radiative processes (photoionization) alone.  These non-photoionizing (shorthand for shocks, etc.) mechanisms, therefore, may contribute an appreciable amount to the excitation of the warm-phase ISM.  In recent studies of non-photoionizing mechanisms in starburst galaxies, \citet{calzetti04} found that 3--4\% of the H$\alpha$ emission in nearby 
starburst galaxies were produced by non-photoionizing sources and that the main cause of this 
non-photoionized emission was due to the mechanical energy injected from the recent and past star formation.  These conclusions are significant because they are one of the first measurements of the small-scale feedback mechanisms from stars (i.e., stellar winds \& supernovae explosions) produced in the starburst.  
Knowing the importance of star formation of the surrounding ISM on small-scales allows us to understand how larger amounts of star formation 
affect the global ISM, and thereby, galaxy evolution.  Such global 
feedback mechanisms can drive the remaining gas, 
necessary to produce stars, from the galaxy to enrich the intergalactic medium 
(IGM).  Such expansion of the ionized gas into the IGM appears to  be occurring in NGC 1569, according to H$\alpha$ 
\citep{heckman95} and X-ray \citep{martin02} studies.  Thus, NGC 1569 is ideal for studying the current non-photoionized processes within the galaxy and 
determining how important these processes are to the future of the ISM composition in this galaxy.  

\subsection{H~{\sc ii} Region Luminosity Function and Size Distribution}

The luminosity function (LF) of H~{\sc ii} regions in a galaxy provides an excellent tracer of the recent (age $\leq$ 10 Myr) star formation.  The form of this function is typically given as a power law in the form:
\begin{equation}
N(L) dL = A L^{- \alpha} dL,
\end{equation}
where $N(L) dL$ is the number of H~{\sc ii} regions with a H$\alpha$ luminosity between $L$ and $L+dL$, $A$ is a constant, and $\alpha$ is the slope of the power law.  
Past studies \citep[e.g.,][]{kennicutt89} have found that the power law decreases for later galactic types (1.75 for LMC, 2.1 for M101, and 2.3 for M31).  Some galaxies show a break in the LF slope where the fainter H~{\sc ii} regions have a shallower slope \citep[e.g., ][]{rozas96}.  
Some suggest that this slope break is due to a transition between normal H~{\sc ii} regions and giant H~{\sc ii} regions \citep[e.g.,][]{kennicutt89}.  \citet{oey98} suggest that the evolution of the ionizing stars in the clusters causes this slope break.  

The H$\alpha$ luminosity function of NGC 1569 has been determined in the past by \citet{youngblood99}.  They find 31 H~{\sc ii} regions and derive a slope of -1.38$\pm$0.16.  They also find that the slope of the luminosity function turns over at lower luminosities.  With the improved spatial resolution of HST, we find a factor of 30 more H~{\sc ii} regions in NGC 1569 than was previously reported and a slope similar to that of \citet{youngblood99}.

The size distribution of H~{\sc ii} regions provides insight into the evolutionary state of star forming regions within a galaxy.  This size distribution has been fitted by an exponential law in the past \citep[e.g.,][]{bergh81}. However, \citet{oey03} suggest that the size distribution is best explained by a power law distribution because the LF and the size distribution are  connected.  For NGC 1569, \citet{youngblood99} fit the size distribution of the H~{\sc ii} regions with an exponential function and derive a characteristic H~{\sc ii} diameter of 34$\pm$3 pc.  

\subsection{NGC 1569}
NGC 1569 is a nearby \citep[D~=~2.2$\pm$0.6 Mpc;][]{israel88} Im galaxy that has been well-studied over the last 20 years.  Two prominent, stellar-like features are located at the center of this galaxy, which are thought to be super-star clusters \citep[SSCs; see][]{prada94} formed in a starburst 2-10 Myr ago \citep{gonzalez97}. 
However, SSC A has been shown to be two stellar clusters superimposed \citep{maoz01}.  
Several studies using optical interference-filter imagery of the ionized gas show evidence of an eruptive event which occurred in the galaxy's past \citep[e.g.,][]{devost97}.
Kinematics of the ionized gas that were studied by \citet{tomita94} showed that the 
expanding gas moves at speeds from 10 to 100 km s$^{-1}$.  
\citet{heckman95} found that the optical filaments at distances of 2 kpc from the 
center of the galaxy are traveling over 200 km~s$^{-1}$.  

The consensus of all these and other past studies is that a starburst occurred approximately 10 Myr ago.  
This event \citep[of unknown origin but probably produced from an interaction with a H~{\sc i} companion,][]{muehle05} produced super star clusters (SSCs) A1, A2, and B.  
Because of the numbers of massive stars and their rapid evolution
(first giving rise to stellar winds and then supernovae), the galaxy underwent a pronounced
kinematical and morphological change, even disruption, during the past several Myr.  
Evidence suggests that some fraction of the supernova ejected material will escape the galaxy and will enrich the IGM \citep{martin02}. 

With the superior angular resolution of the {\it Hubble Space Telescope} over 
ground-based instruments, we determine the current amount of non-photoionized 
H$\alpha$ emission within the galaxy on a 3 pc by 3 pc scale.  Using previous 
results, we determine the mechanisms necessary to produce this non-photoionized H$\alpha$ emission in this galaxy and whether the past star formation can explain the current amount of non-photoionized emission within the galaxy.  We have also generated a new H~{\sc ii} region luminosity function and size distribution.  
The observations and data reduction are presented in \S \ref{obs}. Basic results of our analysis specifying the fraction of non-photoionized gas in NGC 1569 are given in \S \ref{res}, and discussion of our findings is given in \S \ref{dis}.  
Results and Discussion of the H~{\sc ii} region size distribution and luminosity function are given in \S \ref{res2} and \S \ref{dis2} respectively.  In \S \ref{summ} a summary of our findings and concluding remarks are made.  

\section{OBSERVATIONS AND DATA REDUCTIONS \label{obs}}

{\it WFPC2} images of NGC 1569 were taken 1999 September 23 for the Cycle 8 program, GO-8133. 
NGC 1569 was oriented in two of the wide-field chips (WF2 \& WF3) of the camera with a nearby $10^{th}$ 
magnitude star placed out of the field of view (see Figure \ref{compass}).  
The effective plate scale is 0$\farcs$0996 pixel$^{-1}$ (1.06 pc pixel$^{-1}$ at the adopted distance of 2.2 Mpc).  
The GO-8133 images used here were F486N (H$\beta$), F502N ([\ion{O}{3}]), F547M ($\sim$ Str\"{o}mgren $y$), F656N (H$\alpha$), and F673N ([S~{\sc ii}]). 
NGC 1569's heliocentric radial velocity is $-104$~km~s$^{-1}$. The shift of each emission line is $\sim-2$~\AA, and therefore, all emission lines were observed very near the center of the filter transmission curve.  
Observational parameters for these data are found in Table \ref{tab1}.  
See \citet{buckalew00} for more details on the calibration processes used for these images. 
In that paper, the reduction process was detailed for the F502N image only.  
However, the reduction process is similar for the F486N (H$\beta$), F502N ([\ion{O}{3}]), F656N (H$\alpha$), and F673N ([S~{\sc ii}]) images.  
For the analysis of the non-photoionized gas fractions, the images were binned in 3 pixels $\times$ 3 pixels (3$\times$3), 5 pixels $\times$ 5 pixels (5$\times$5), 7 pixels $\times$ 7 pixels (7$\times$7), and 9 pixels $\times$ 9 pixels (9$\times$9).  Use of these binned images reduced the uncertainties in the emission-line ratio maps.  For the analysis of the H~{\sc ii} region luminosity function and size distribution, a calibrated, unbinned H$\alpha$ image was used.  

The filter transmission curve of F656N covers H$\alpha$ as well as [N~{\sc ii}] $\lambda\lambda$6548,6583.  To remove 
this contamination of the H$\alpha$ emission, we compared the eight values of [N~{\sc ii}]/H$\alpha$ and [N~{\sc 
ii}]/[S~{\sc ii}] from \citet{buckalew05} to determine which varied the least.  \citet{calzetti04} determined that [N~{\sc ii}]/[S~{\sc ii}] varies little over similar starburst galaxies. However, we found that 
[N~{\sc ii}]/H$\alpha$ varies less in NGC 1569 than [N~{\sc ii}]/[S~{\sc ii}].  
Taking into account the blueshift, the wavelengths have changed by 
2 \AA~for the 3 emission lines.  The total system throughput ratios of $\lambda$6548/$\lambda$6563 and $\lambda$6583/$\lambda$6563 were determined, and the total percentage of [N~{\sc ii}] contamination is found to be 1\%.  
The H$\alpha$ image was multiplied by 0.99 to reflect this amount of contamination.  


Flux calibration of the continuum-subtracted, interference-filter images was done using 
the procedure from \citet{buckalew00}.  The flux conversion numbers are in units of erg cm$^{-2}$ DN$^{-1}$ and are 
$1.35\times10^{-14}$ for F487N, $1.07\times10^{-14}$ for F502N, $4.14\times10^{-15}$ for F656N, and $3.95\times10^{-15}$ 
for F673N.  


The image with the lowest signal-to-noise, H$\beta$, was used to generate a mask. We determined the standard deviation of the background in the H$\beta$ image at several sky locations.  The standard deviation was 2.5$\times$10$^{-17}$ erg s$^{-1}$ cm$^{-2}$ in the 3$\times$3 H$\beta$ image.  Standard deviations were also measured for the 5$\times$5, 7$\times$7, and 9$\times$9 binned images. We generated H$\beta$ masks at the 3$\sigma$, 5$\sigma$, 7$\sigma$, and 9$\sigma$ levels above the mean sky to multiply to the appropriate binned images.  The results discussed below will focus predominantly on the 3$\times$3 binned images that have been masked to remove pixels with a flux level less than 5$\sigma$ above the mean sky (hereafter 3$\times$3 5$\sigma$).


Three ratio maps were made of the galaxy: H$\alpha$/H$\beta$, [O~{\sc iii}]/H$\beta$, and [S~{\sc 
ii}]/H$\alpha$.  
The H$\alpha$/H$\beta$ ratio map was used to deredden the [O~{\sc iii}]/H$\beta$ and [S~{\sc ii}]/H$\alpha$ maps.  
The pixels in the H$\alpha$/H$\beta$ ratio map with H$\alpha$/H$\beta$ $<$ 2.86 were removed because 2.86 is our assumed theoretical ratio of these 
two hydrogen lines \citep{osterbrock89}, typical for a H~{\sc ii} region with T$_{\rm e}$ of 
10,000 K and N$_{\rm e}$ of 100 e$^-$ cm$^{-3}$.  The following equation was used to convert H$\alpha$/H$\beta$ to C(H$\beta$): 
\begin{equation}
C(H\beta) = (log[H\alpha/H\beta]-log[2.86])/0.326, 
\end{equation}
where H$\alpha$/H$\beta$ indicates a value from our ratio map.  
With the C(H$\beta$) ratio map, we dereddened the other ratio maps using the following formulae:  
\begin{equation}
I([O~III])/I(H\beta) = F([O~III])/F(H\beta)\times10^{-0.03C(H\beta)}, 
\end{equation}
\begin{equation}
I([S~II])/I(H\alpha) = F([S~II])/F(H\alpha)\times10^{-0.02C(H\beta)}, 
\end{equation}
where I() means the dereddened fluxes and F() indicates the reddened fluxes from the ratio maps.  The multiplicative factors to C(H$\beta$) are based on the reddening law from \citet{seaton79}.  The average C(H$\beta$) for this 
galaxy is 0.69 (E(B-V) = 0.47 \& A$_V$ = 1.5 mag).  Thus, the reddening only decreases the ratios by an average of 5\%.  
With the dereddened ratio maps, we determined which 3$\times$3 5$\sigma$ [O~{\sc iii}]/H$\beta$ and [S~{\sc ii}]/H$\alpha$ ratios
had a reasonable H$\alpha$/H$\beta$ ratio (i.e.\ H$\alpha$/H$\beta>2.86$).  
Figure \ref{o3hb} displays the 3$\times$3 5$\sigma$ [O~{\sc iii}]/H$\beta$ ratio map, and Figure \ref{s2ha} displays the 3$\times$3 5$\sigma$ [S~{\sc ii}]/H$\alpha$ ratio map.  

\section{EMISSION MECHANISM RESULTS \label{res}}

\subsection{Determination of Photoionized and Non-Photoionized Regions \label{nonpiresults}}

In Figure \ref{lo3hbvls2ha} we plot the 3$\times$3 5$\sigma$ [O~{\sc iii}]/H$\beta$ versus [S~{\sc ii}]/H$\alpha$ values to compare with the maximum starburst relation from \citet{kewley01}.  This line denotes the upper limit between photoionized gas (below the line in Figure \ref{lo3hbvls2ha}) and ionized gas from shocks or other non-photoionizing sources (above the line in Figure \ref{lo3hbvls2ha}).  
All points above this maximum starburst line are counted as emission due to non-photoionizing processes.  
All points below the maximum starburst line are counted as due to photoionizing processes.  
The distribution of the NGC 1569 points is similar to that of NGC 4214 \citep{calzetti04}.  
From Figure \ref{lo3hbvls2ha} we find that 16\% of the total number of points fall above the maximum starburst line. 
This areal percentage is similar to those found for NGC 3077, NGC 4214, 
NGC 5236, and NGC 5253 \citep{calzetti04}.  The ratio of the H$\alpha$ luminosity in  these non-photoionized points to the total 3$\times$3 5$\sigma$ H$\alpha$ luminosity is 15\%.   

Compared to previous results from \citet{calzetti04}, this percentage is four to five times larger than the percentages of NGC 3077, NGC 4214, NGC 5236, 
and NGC 5253 (range 3--4\%).  
Note that the method for determining non-photoionized results is different between 
ours and that of \citet{calzetti04}. We choose everything above the maximum 
starburst line while \citet{calzetti04} use an additional criterion (their dashed
line in their Figure 7).  This dahsed line, in a basic way, limits 
log([S{\sc ii}]/H$\alpha$) to values $\geq$-0.5 or $\geq$-0.4.  
These differences in selection criterion are based on differences in the shock 
models used.  We use the shock models of \citet{hartigan87}; 
see \citet{calzetti04} for their shock models.  
If we use their additional limitation, we determine a non-photoionized percentage
of 2--1\%. Thus, we can simply say that the non-photoionized percentage of emission
is different in NGC 1569 than that of other irregular galaxies regardless of the 
shock models used.   

The 3$\times$3 3$\sigma$, 3$\times$3 7$\sigma$, and 3$\times$3 9$\sigma$ non-photoionized H$\alpha$ luminosity percentages of NGC 1569 are 19\%, 12\%, and 10\% 
respectively.  
The limited decrease in non-photoionized percentage as a function of increasing $\sigma$ suggests that the non-photoionized emission of NGC 1569 is found in high surface-brightness regions.  This lack of decrease is at odds with what \citet{calzetti04} found for other, similar irregular galaxies.  The fraction of non-photoionized flux found in the low signal-to-noise threshold (5$\sigma$) maps should be much larger (not just a factor of 1.5) than the fraction found in the higher signal-to-noise threshold maps.  The lower signal-to-noise ratio maps should give a higher percentage simply because more lower surface brightness regions would be counted, typically the location of non-photoionized emission.  We discuss in \S \ref{dis} possible sources to explain this high surface brightness non-photoionized emission in NGC 1569.  

Because the percentage found from the 3$\times$3 5$\sigma$ maps is so much larger than the previous results, we determined the fraction of non-photoionized H$\alpha$ emission from the 5$\times$5, 7$\times$7, and 9$\times$9 5$\sigma$ maps.  
The 5$\sigma$ non-photoionized H$\alpha$ luminosity percentages for 5$\times$5, 7$\times$7, and 9$\times$9 are 16\%, 10\%, and 10\% respectively. 
Increasing either the bin size to 9$\times$9 or the sigma to 9$\sigma$ causes a 5\% decrease in the non-photoionized H$\alpha$ luminosity.  
The 9$\times$9 9$\sigma$ result is 5\%, a factor of 3 decrease, but still the largest percentage for an irregular galaxy.  The majority of these results fall at 10\%$\pm$3\%, which is the percentage that we claim for NGC 1569, whether we use 3$\times$3 5$\sigma$ or 9$\times$9 9$\sigma$ maps.  The decrease in the fraction of non-photoionized luminosity with bin size is caused by spatial dilution with the photoionized emission.  

The idea of spatial dilution brings up the important point that NGC 1569 is closer, by a factor of 2, than the irregular galaxies studied in \citet{calzetti04}.  Since they use a 3$\times$3 5$\sigma$ result, we must compare a similar spatial result to theirs.  The results above that are comparable in spatial resolution are the 5$\times$5 5$\sigma$ and 7$\times$7 5$\sigma$ results (the best being a 6$\times$6 5$\sigma$ result).  Both of these results are 10\%.  Thus, our final, accepted percentage of 10\%$\pm$3\% works well for comparison with these further irregular galaxies.  

\subsection{Further Quantities Derived from the H$\alpha$ Images and Ratio Maps}

From the H$\alpha$ luminosity of the 3$\times$3 5$\sigma$ results, we derive the star formation rate (SFR) and number of hydrogen ionizing photons per second (Q$_0$) using the equations from \citet{kennicutt98}.  These results, along with those from \S \ref{nonpiresults}, are summarized in Table \ref{tab2}.  When we compare these results to those of similar nearby starburst galaxies in \citet{calzetti04}, we find that the SFR and Q$_0$ of NGC 1569 are similar.  This similarity is also true of the total H$\alpha$ luminosities and areal coverage of the non-photoionizing luminosity.  However, we showed in \S \ref{nonpiresults} that the major difference is the ratio of non-photoionized H$\alpha$ luminosity to the total luminosity, which we will discuss further in \S \ref{dis}.  

We have compared the properties of the photoionized and non-photoionized properties 
using a Wilcoxon test to determine differences between the samples.  
As confirmation that the Wilcoxon test works, 
we find that [O~{\sc iii}]/H$\beta$ and [S~{\sc ii}]/H$\alpha$ are much higher in non-photoionized areas compared to photoionized areas.  
We find that the southern areas of NGC 1569 are prevalently non-photoionized areas.  
The visible outflow and H$\alpha$ shell lie on this side of NGC 1569.  
This outflow lies on the near side of NGC 1569, making the non-photoionized emission more prominent.  
We also find that non-photoionized areas tend to have larger C(H$\beta$), which can be explained by dust or non-photoionized mechanisms.  
\citet{hartigan87} shows that shock excitation can elevate H$\alpha$/H$\beta$. 
Finally, we find that the continuum flux is higher at the locations of the non-photoionized emission compared to the locations of the photoionized emission.  Perhaps, this finding means that 
the non-photoionized emission is typically coming from areas closer to stars and star clusters, but this interpretation is speculative at best.  What we find below
is that while some non-photoionized emission is near bright continuum sources, the majority is scattered throughout the galaxy neither tending to be near or far from resolved stars and star clusters.  

\section{DISCUSSIOIN OF NON-PHOTOIONIZED EMISSION \label{dis}}

A significant fraction of the H$\alpha$ luminosity in this galaxy arises from non-photoionizing sources.  NGC 1569 is riddled with H$\alpha$ shells, arcs, and filaments throughout the main body of the galaxy (see Figure \ref{compass}). Can the mechanical energy from the stellar winds and supernovae explain the amount of non-photoionized H$\alpha$ luminosity? Do potential sources such as Wolf-Rayet stars and supernovae remnants or different gas phases, such as the X-ray hot gas, explain the locations of some of these non-photoionized sources?   

\subsection{Is the Starburst Enough?}
\subsubsection{Continuous Star Formation}

We derive a SFR of 0.15 M$_{\odot}$ yr$^{-1}$ from the H$\alpha$ luminosity.  We 
generated Starburst 99 \citep{leitherer99} continuous star formation models from 0 Myr to 100 Myr to determine the amount of mechanical luminosity generated over this time frame.  The model parameters are the metallicity of Z = 0.004, lower and upper stellar mass limits of 0.1 M$_{\odot}$ and 100 M$_{\odot}$, and a Salpeter initial mass function (IMF).  The 10 Myr and 100 Myr model mechanical luminosities are 3.02$\times$10$^{40}$ ergs s$^{-1}$ and 8.91$\times$10$^{40}$ ergs s$^{-1}$ respectively.  We follow the prescription of \citet{calzetti04} and assume that the H$\alpha$ luminosity due to the mechanical luminosity is 2.5\%.  
The H$\alpha$ luminosity due to these mechanical energy sources is 7.5$\times$10$^{38}$ ergs s$^{-1}$ and 
2.2$\times$10$^{39}$ ergs s$^{-1}$ for 10 Myr and 100 Myr respectively.  Our estimate of the non-photoionized H$\alpha$ luminosity is 2.9$\times$10$^{39}$ ergs s$^{-1}$.  These model non-photoionized H$\alpha$ luminosities are 4--1.3 times too low to explain the current amount of non-photoionized H$\alpha$ luminosity that we estimate.  
\citet{greggio98} state that $\sim$5 Myr ago the SFR was 0.5 M$_{\odot}$ yr$^{-1}$ and that this SFR was constant over the past 100 Myr.  If we assume that the SFR was 0.15 M$_{\odot}$ yr$^{-1}$ over the past 5 Myr, the mechanical luminosity of this contribution is 10 times too low to explain the non-photoionized H$\alpha$ luminosity.  The SFR of 0.5 M$_{\odot}$ yr$^{-1}$ can provide a sufficient amount of mechanical luminosity after $\sim$10-20 Myr to explain the current non-photoionized H$\alpha$ luminosity seen.  
We agree with \citet{greggio98} in so far that the SFR must have been higher in the past to explain the amount of non-photoionized H$\alpha$ emission currently seen, if this emission is from the mechanical energy of supernovae and stellar winds.  


\subsubsection{Instantaneous Star Formation}

We assume that the star formation occurred instantaneously and that the stellar mass produced in the burst is equivalent to the most massive star clusters.  According to \citet{anders04}, the mass of the three largest clusters is between 6.2--6.6 log(M$_{\odot}$); similar masses were found in \citet{buckalew05}.  We used the \citet{anders04} mass range in an instantaneous burst Starburst 99 model (other parameters are similar to those used in the continuous star formation) to determine the lower and upper mechanical luminosities at 10 Myr and 100 Myr, which were 
converted to H$\alpha$ luminosities using the same assumptions as before.  The H$\alpha$ mechanical luminosity range for 
10 Myr and 100 Myr is 7.1$\times10^{38-39}$ ergs s$^{-1}$ and 1.6$\times10^{32-33}$ ergs s$^{-1}$ respectively.  At 100 Myr, the non-photoionized H$\alpha$ luminosity remaining long after the burst is minimal. But at 10 Myr, the non-photoionized H$\alpha$ luminosity is similar to that found in our study, and a majority of the mass range gives enough non-photoionized H$\alpha$ luminosity to explain our results as well.   This result is consistent with independent measurements of the star formation history that indicates that the burst probably occurred $\sim$10 Myr ago \citep[e.g.,][]{greggio98}. It also suggests that the SSCs and bright H~{\sc ii} region are responsible for the majority, if not all, non-photoionized H$\alpha$ emission within this galaxy.  However, not all non-photoionized H$\alpha$ luminosity is necessarily explained by these sources. Some non-photoionized emission may be explained by individual events/sources.

\subsection{Sources Responsible for the Morphology of the Non-Photoionized Gas \label{responsible}}

We have already shown that the past star formation can explain the amount of non-photoionized gas.  However, we can 
explicitly show that certain types of objects (e.g.\ Wolf-Rayet stars) are responsible for specific non-photoionized emission and the morphology of that emission.  We compare the locations of the non-photoionized H$\alpha$ with the locations of detected radio supernovae, Wolf-Rayet stars, star clusters, radio point sources, X-ray point sources, X-ray extended emission, and CO molecular cloud detections. 

\subsubsection{Wolf-Rayet Stars \& Star Clusters with Wolf-Rayet Stars \label{wolfrayet}}
Wolf-Rayet stars make good sources of mechanical ionization because they have strong winds. They are also indicators that star formation has occurred recently.  In Figure \ref{n1569wr}, we plot the positions of Wolf-Rayet stars from \citet{buckalew00} with the non-photoionized points on the H$\alpha$ image.  The positions of the Wolf-Rayet stars are 2 pixel-wide black squares and labeled with their name from \citet{buckalew00}.  The non-photoionized points are 1 pixel-wide white squares.  

Wolf-Rayet star S1 has a few non-photoionized points adjacent to its location.   
Wolf-Rayet stars S2 through S5 and Unknown He~{\sc ii} sources U1 through U3 have very little non-photoionized gas surrounding them.
An arc lies above and 
to the left of S5 in Figure \ref{n1569wr}, but we would not expect this particular arc to be caused by S5. SSC A (C4) is probably the main cause of this arc.  Wolf-Rayet star S6 shows 3 non-photoionized regions adjacent to or on top of its location.  
Wolf-Rayet star S7 has no associated non-photoionized emission.  This 
non-detection may result from low signal-to-noise in the H$\beta$ image, or the possibility exists that the wind from S7 is too slow to produce significant mechanical luminosity.

In the star clusters with Wolf-Rayet stars, cluster C1 has a few non-photoionized points adjacent to it.  \citet{buckalew00} showed 
that this ``Wolf-Rayet" star cluster also had extended He~{\sc ii} emission, which is coincident with the non-photoionized points.  This extended 
He~{\sc ii} emission signifies a shock velocity of $\sim$120 km s$^{-1}$, the shock velocity where a maximum of He~{\sc ii} emission 
is produced \citep{garnett91}.  The wall of 
non-photoionized points between C1 and C3 could be due to a wind-wind interaction between these clusters.  
C2 has one point of shocked gas.  This non-photoionized point lies in the same direction as the extended He~{\sc ii} emission found in \citet{buckalew00}, implying a shock velocity of $\sim$120 km s$^{-1}$.  
C3 has a significant amount of non-photoionized gas on top of and around this star cluster.  
The line of points running through it do not trace any noticeable arc or filament in the H$\alpha$ image (Figure 
\ref{n1569wr}).  
C4 (aka SSC A) is adjacent to S5 and the non-photoionized arc near C4 is probably caused by stellar winds from C4.
C5 has an arc of non-photoionized emission to its lower right.  This young cluster is probably the culprit for the majority of this non-photoionized emission.  However, the massive star clusters in that H~{\sc ii} region may also be contributing significantly, possibly this is a wind-wind interaction between the two star clusters.  
The average distance between the line and cluster 
is $\sim$6 pc.  
If we assume a shock velocity of 100 km s$^{-1}$, then the shocked gas originated a minimum of 10$^{4-5}$ yrs ago.  
Given the brevity of the Wolf-Rayet phase, this time scale would suggest that the onset of the Wolf-Rayet phase is responsible for the line of non-photoionized emission.

\subsubsection{Star Clusters \label{clusters}}

We have plotted the WFPC2 F555W images of GO-6111 \citep{greggio98} and GO-6423 \citep{hunter00} 
along with the non-photoionized points in Figure \ref{n1569cluster}.  The positions of important star clusters are designated by their name from \citet{hunter00} and a line connects that name to the approrpiate star cluster.  The non-photoionized points are 1 pixel-wide white squares.
Four arcs of non-photoionized gas near SSC A and SSC B are very prominent.  The arc closest to SSC A appears to be 
wrapped around a cluster/star south of SSC A.  Perhaps the wind of this star is interacting with a nearby, denser ISM.  
However, this arc is apparent in the left panel of Figure \ref{n1569cluster} but not the right panel.  This arc (although more a line) 
is apparent in Figure \ref{hanonpi} and must not, therefore, be an artifact due to the coordinate transformation.  The 
cluster/star is the nearest stellar object inside this arc.  No other point source or extended source can 
explain this arc.  The other arc is just west of SSC B.  This arc is apparent in Figures \ref{hanonpi} and \ref{n1569cluster} and could be due to a 
wind-wind interaction between SSC B and cluster 28 (designation from Hunter et al.\ 2000) 
and the other cluster/star south of 28 that has no designation. The last arc is west of SSC A and is associated 
with cluster 10 and possibly clusters 13, 14, \& 15.  The arc is between cluster 10 (a.k.a.\ C1 from \S \ref{wolfrayet}) and clusters 13-15, again 
a possible wind-wind interaction between these clusters.  
Another arc is present and associated with clusters 39 and 40.  Cluster 39 is  C5 from \S \ref{wolfrayet} and has 
already been discussed.  Clusters 30 and 35 have non-photoionized emission associated with their position (see Figure \ref{n1569cluster}). Clusters 29 and 34 have at least one non-photoionized point near their 
location and are bright X-ray sources that are mentioned in \S \ref{x}.
Finally, a large shell of non-photoionized gas surrounds the clusters 6, 7, 9, and 10.  This shell is supported by the 
winds of these clusters and extended X-ray emission (discussed in \S \ref{x}).  Several other sources such as the thermal radio source M-1 from \citet{greve02} and the molecular cloud 3 
from \citet{taylor99} are also present at this region and help to show that this location is currently a site of 
vigorous star formation. 

\subsubsection{Radio Sources from \citet{greve02} \label{snr}}

In Figure \ref{n1569snr} we plot the radio-detected sources, such as supernovae and supernova remnants (SNRs), from 
\cite{greve02} along with the non-photoionized locations.  The positions of the radio sources 
are 2 pixel-wide black squares and labeled with their name from \citet{greve02}.  The non-photoionized points are 
1 pixel-wide white squares.  The grayscale scheme used in \S \ref{wolfrayet} is implemented 
here.  
Note that radio sources M-a through 
M-d are tentative detections and are not known to be radio supernovae or SNRs yet.  

M-1 and M-b,c,d are found inside a large shell of non-photoionized points.  These points surround the brightest H~{\sc 
ii} region complex within the galaxy.  M-1, a thermal source according to \citet{greve02}, is probably just the thermal radio 
emission from this large H~{\sc ii} complex termed region 2 in \citet{waller91}.  
M-2 is a radio supernovae or SNR with one non-photoionized point adjacent to our marker. 
M-3 is a radio supernovae or SNR according to \citet{greve02}.  This source is situated in the middle of a large group of 
non-photoionized points.  M-3 is more than likely responsible for some of these.  However, we show in the next section 
that a X-ray binary is also found in the same area.  
According to \citet{greve02}, M-4 is a thermal source (i.e., the H~{\sc ii} region found below the point marker), and a few non-photoionized points lie  near the marker.  
M-5 is a thermal source, according to \citet{greve02} and is probably associated with the coincident H~{\sc ii} region.  
However, a H$\alpha$ shell is found at the upper right hand corner of this H~{\sc ii} region and is the source of the large number of 
non-photoionized points to the upper right of M-5.  We think that M-5 is this shell and not the H~{\sc ii} region. 
M-6 is a SNR according to \citet{greve02} and was also detected optically by \citet{shopbell00}.  As 
a side note, the size of the SNR in the radio is smaller than the optical counterpart (17 pc versus 21 pc).  
The brightest section of the SNR is coincident with the non-photoionized points.  
VLA-8 is most likely a SNR as well as VLA-16 \citep{greve02}.  Regardless of their definition, these objects appear to contribute to the non-photoionized emission in their surroundings.  The only SNR or radio supernovae  without any nearby non-photoionized gas is VLA-11, probably caused by the lack of signal-to-noise in our H$\alpha$ image at the position of VLA-11.

\subsubsection{X-ray Point Sources \label{x}}

In Figure \ref{n1569x} we plot the X-ray point sources from \citet{martin02} along with the non-photoionized locations on the H$\alpha$ image.  
The positions of the X-ray sources 
are 2 pixel-wide black squares.  The non-photoionized points are 
1 pixel-wide white squares.
We label each \citet{martin02} source with their number from that article along with a one letter identification indicating the nature of the source: C for cluster, S for SNR, and X for X-ray binary.  
Objects 19C and 22C are clusters 29 and 34 from \citet{hunter00}.  Both clusters have at least one non-photoionized 
point adjacent to their position.  More than likely, these points are associated with those clusters, especially 19C since no other object exists in this evacuated cavity.  
\citet{martin02} were unsure if 28S was indeed a SNR.  Comparing the \citet{greve02} results to these, we find 
that 28S is M-6, which is also the optical SNR discovered by \citet{shopbell00}.  
We discussed the non-photoionized gas around this point in \S \ref{snr}.  
All X-ray binaries except for 21X, 25X, 26X, and 29X have some nearby non-photoionized gas.  
Of these without non-photoionized gas, 21X and 26X sit inside large shells which have shocked gas along their edges.  
Perhaps these two X-ray binaries or their predecessors are responsible for these shells.  
Two of the remaining X-ray binaries deserve comment.  
14X lies near M-3, the SNR.  Thus, the SNR may not be completely responsible for this large 
non-photoionized area, or 14X was misdiagnosed as a X-ray binary and is actually the X-ray detection of M-3.  
Similarly, 24X sits atop the line previously explained by winds from C5, a cluster with detected Wolf-Rayet stars, 
and is possibly responsible for some percentage of the mechanical H$\alpha$ luminosity found at this location.

\subsubsection{Extended X-ray Emission \label{chandra}}

In Figure \ref{n1569chandra}, we show the 0.3--7 keV X-ray image from \citet{martin02} along with the locations of the non-photoionized gas.
One correlation is seen between the extended emission and non-photoionized gas, 
the large X-ray extended emission located over the brightest H~{\sc ii} region complex (marked with a black circle in Figure \ref{n1569chandra}). 
This location was also detected as M-1 in \citet{greve02}, a thermal source attributed to the H~{\sc ii} region complex termed 2 in \citet{waller91}.

\subsubsection{CO Emission from Molecular Clouds \label{co}}

We have plotted in Figure \ref{n1569co} the non-photoionized points on a CO map from \citet{taylor99}. 
The different molecular clouds are labeled using the designations from \citet{taylor99}.  
The only clouds found near non-photoionizing sources are 3 and 4.  
The molecular cloud 3 is found in the extended X-ray and thermal radio emission associated with the brightest H~{\sc ii} region complex.  
The cloud appears to be circumscribed by the non-photoionized gas.  
Also, the local minimum in CO emission between 3 and 1+2 is coincident with one side of the non-photoionized shell circumscribing 3.  
The molecular cloud 4 is found coincident with some non-photoionized gas but too few non-photoionized points are present in 
that location to define the morphology of the cloud.  
Other low level features not labeled on the CO map are noise.

\subsubsection{Summary}


The most important correlation of the non-photoionized points with individual sources is with the largest H~{\sc ii} complex.  
A ring of non-photoionized emission encircles the H~{\sc ii} region complex 2 \citet{waller91}.  This complex is the largest site of current star
 formation in NGC 1569.  This site is also host to a large thermal radio source, extended X-ray emission, molecular cloud cores, supernovae, Wolf-Rayet
 stars, and star clusters. The stellar winds and supernovae explosions caused by this vigorous star formation are causing a large evacuation of gas which is striking the nearby ISM.  Future kinematical evidence may show that the velocity of the gas is greater than the escape velocity.  If it will escape, this complex may be similar (yet smaller) to the H~{\sc ii} complex that contained SSCs A1/2 \& B.  In a few Myr, NGC 1569 may have another large H$\alpha$ arm similar to the one currently ejected due to the super star clusters' creation.  

Other features of the non-photoionized emission appear to be related to SSCs A1/2 \& B.  A few arcs of non-photoionized 
emission seem to be near these two clusters and could be attributed to stellar winds emanating from these clusters and interacting with the winds from
other nearby star clusters.  Several other star clusters with Wolf-Rayet stars, without Wolf-Rayet stars, with X-ray emission, and/or with thermal radio 
emission appear with non-photoionized emission on top of or adjacent to these systems.  The most important is C5 \citep{buckalew00}.  A significant arc 
of non-photoionized emission appears around this star cluster with Wolf-Rayet stars.  The onset of the Wolf-Rayet phase in the OB stars is the likeliest 
explanation for the shocked emission.  

Other point sources that have non-photoionized emission are radio and optical supernovae, low-mass X-ray binaries, and Wolf-Rayet stars.  With some of 
Wolf-Rayet stars, we notice that the position of the non-photoionized emission is in the same direction as the extended He~{\sc ii} emission emanating 
from these sources.  We suggest that this indicates the shock velocity is $\sim$120 km s$^{-1}$, the velocity where He~{\sc ii} 4686 \AA~emission 
peaks.  

\subsection{Origin of the large percentage of non-photoionized gas in NGC 1569}

\citet{calzetti04} stated that the maximum percentage of non-photoionized H$\alpha$ luminosity should be no more than 10--20\%.
NGC 1569 lies within this range.  These maximum percentages occur when the number of supernovae or stellar winds shocking the ISM stays the same and the level of photoionized emission is decreasing (i.e., when the OB stars are exploding as supernovae).  This scenario typically occurs for older ages of a starburst, 6--20 Myr roughly.  
Previous papers \citep[e.g.,][]{gonzalez97} concluded that NGC 1569 is in a post starburst phase.  
The ages of the super star clusters or large star clusters in NGC 1569 have ages around 10-20 Myr.  
Thus, the major star formation event is over, but star formation still occurs in smaller areas of the galaxies.  
The shocks produced from the major star formation event, which produced SSCs A1, A2, and B, are producing the non-photoionized H$\alpha$ emission while a new, smaller (by a factor of 3) amount of star formation provides the photoionized component.  

\section{MEASURING THE H II SIZE DISTRIBUTION AND LUMINOSITY FUNCTIONS \label{res2}}

We employed HIIphot \citep{thilker00} to determine the positions, sizes, and fluxes of H~{\sc ii} regions in NGC 1569.  
We used the values of 0.75, 1, 1.25, 1.5, 2, 3, and 10 for the input terminal gradient parameter that determines when the code stops determining the boundary of the H~{\sc ii} region.  
The program generated $\sim$6500 potential H~{\sc ii} regions over the entire galaxy.  
Because the galaxy has a tremendous filamentary structure, the program interpreted each arc as a separate H~{\sc ii} region.  
To remove the detections of the filamentary structures as H~{\sc ii} regions, we accepted only those sources with a signal-to-noise of 100 or 
greater.  
After visual inspection to certify that the list does not contain filamentary structures, the actual number of H~{\sc ii} regions used in the analysis was reduced to 1018.
To deredden the luminosities of each source, we took the positions of the H~{\sc ii} regions and calculated the mean 3$\times$3 
5$\sigma$ H$\alpha$/H$\beta$ value at these positions. 
The aperture used to calculate this mean was equal to the size of the H~{\sc ii} region.  
In some cases, no valid points were found around the H~{\sc ii} region, and we accepted 
a C(H$\beta$) value equal to the average photoionized C(H$\beta$) of the 3$\times$3 5$\sigma$ results (0.67).  
The final range of H$\alpha$ luminosities of the H~{\sc ii} regions was 10$^{36.4}$--10$^{38.91}$ ergs s$^{-1}$.  
The effective FWHM measurements of this final luminosity sample of H~{\sc ii} sources were converted to 3$\sigma$ H~{\sc ii} region diameters, producing a range of 6.3--44.8 pc.  

\subsection{Computing the H~{\sc ii} Region Size Distributions \label{computesd}}

A diameter histogram of the $\sim$1000 H~{\sc ii} regions was generated.  The size of the bins are dictated by the size of the sample and by the standard deviation of the sample.  The equation relating these three values is given by the equation of \citet{scott79}:
\begin{equation}
Bin size = 3.5 \sigma n^{-1/3}, 
\end{equation}
where $\sigma$ is the standard deviation and $n$ is the sample size.  
These bins stretched to include the minimum value at the far left of the first bin 
and were added until the entire sample was covered.  
Uncertainties for each bin were estimated by assuming that the uncertainties of the bin were Poissonian (i.e.\ $\sqrt{n}$).  
Because we plot the log of the bin sizes in Figure \ref{h2sd}, 
the uncertainties of each bin are ($\sqrt{n}$ln[10])$^{-1}$.  

Using the frequency values and uncertainties, we assumed that the H~{\sc ii} region size distribution followed a power law like \citet{oey03}.
Linear regression was used to determine the slope (i.e.\ power law index) with the logarithmic values of the frequency (i.e.\ 
number of points in a bin) and diameter.
The first fit taking into account all bins was significant to 23\%, which is not a statistically significant fit.  
Removing the aberrant second left value from the fit, we derive a marginally significant (10\%) fit to the data.  
The equation is the following:
\begin{equation}
log(N) = -3.02\pm0.27 \times log(D) + 5.16\pm0.27,
\end{equation}
where N is the frequency of a certain bin and D is the diameter value of the middle of the bin. 
Others \citep[e.g.,][]{youngblood99} have fit the size distribution with an exponential function.  We have also fit the size distribution with an exponential function.  The significance of such a fit (15\%) is lower than the power law fit.  For completeness, the equation is 
\begin{equation}
N = e^{-0.23\pm0.02 \times D + 7.54\pm0.23},
\end{equation}
where the letters represent the same values as the previous equation.

\subsection{Computing the H~{\sc ii} Region Luminosity Functions \label{computelf}}

Using the logarithmic H$\alpha$ luminosities of the H~{\sc ii} regions, 
we generated a luminosity histogram.  
We follow the procedure outlined in the previous section for generating histograms.  This histogram, displayed in Figure \ref{h2lf}, was 
fitted by a power law distribution.  
The 3 left-most points are found below the fourth.  
The fit of all points except the four leftmost is significant to 8\%:  
\begin{equation}
log(N) = -1.00\pm0.08 \times log(L) + 39.25\pm2.14, 
\end{equation} 
where N is the number of points in one bin and L is the average luminosity for the bin.  
For comparison, the slope of the luminosity function for 
NGC 1569 in \citet{youngblood99} is -1.38$\pm$0.16, 
which is not within the 3$\sigma$ confidence interval of our result. 
The luminosity range of \citet{youngblood99} 
is between Log(L)$=$38.0--39.9.  
Our fitted logarithmic luminosity range is 36.4--38.91.
We have plotted the slope of \citet{youngblood99} in 
Figure \ref{h2lf} over the luminosity range of their data.  
Note that their slope does work for our points  
found in a similar luminosity range.  
The most significant fit is for the luminosity range 37.22--38.91 and is the following:
\begin{equation}
log(N) = -1.22\pm0.06 \times log(L) + 47.504\pm2.14.
\end{equation}
This slope is found within the 3$\sigma$ confidence intervals of the slope from \citet{youngblood99} 
and vice versa.  
The best fit equations along with the correlation coefficients for both the  
size distribution and luminosity function are given in Table \ref{tab4}.  

Because of the turnover of the luminosity function at the logarithmic 
luminosity value $\sim$37.0, incompleteness corrections
may be necessary.  We used IRAF to add elliptical galaxies to the image for HIIPHOT to detect.  Elliptical galaxies were used because
they are more extended than point sources and because no random H~{\sc ii} region maker exists. 

The luminosity range of elliptical galaxies ranged from logarithmic values of 36.4 to 37.2.  Above the logarithmic value of 37.1, HIIPHOT detects 100\%
of the added ``H~{\sc ii}" regions.  Below this value, HIIPHOT fails to detect 8\% of the added H~{\sc ii} regions (incompleteness corrected bins are
displayed in Figure \ref{h2lf}).  Adding this correction factor to
those luminosity bins produces little change in the slope value.  The slope value for the incompleteness corrected bins produces a slope change of 0.01.  This small change is within the confidence interval of the original slope.  Thus, the contribution of the 
incompleteness corrections is insignificant.  With this incompleteness method, nothing can be said about corrections to the size distribution.  The added sources are not H~{\sc ii} regions but elliptical galaxies with similar H$\alpha$ luminosities.  

We also produced luminosity functions and size distributions of three subsamples of the H~{\sc ii} regions (see Figure \ref{f13}).  These three 
subsamples were determined based on their 
position from the ``center of the starburst."  
We place the center of the starburst at the location halfway between SSCs A(1 and 2) and B.  The location of all H~{\sc ii} regions in our sample have a minimum distance of 4 pc (maximum of 630 pc), 
suggesting that the winds and shocks have effectively terminated star
formation in this small cavity.  
The first annuli is 115 pc wide with the second is 62.7 pc wide, indicating that massive amounts of star formation are evacuated from the super star cluster area but are occuring rapidly outside this area.  
The luminosity function and size distribution fits for these subsamples were
all significant, but the slopes for the size distribution and luminosity function are the same regardless of position from the super star clusters.
The H~{\sc ii} region surface density was computed for each annuli, and these densities within the annuli increase by a factor of 1.33 from the first to the second annulus.  
These results indicate that these feedback effects in NGC~1569 are confined to the
immediate vicinity of the most recent massive star formation event on
scales of $\sim1$ pc.


\section{DISCUSSION OF H II REGION SIZE DISTRIBUTION AND LUMINOSITY FUNCTION \label{dis2}}

\subsection{H~{\sc ii} Region Size Distribution}

Our power law slope of -3.02 is smaller than slopes of 
similar irregular galaxies \citep[-3.39 to -4.16;][]{oey03}.  
\citet{oey03} state that a power law slope would provide only a good description of the size distribution for 
H~{\sc ii} regions greater than 130 pc in diameter and 
that a slope should be zero for diameters less than 130 pc.  
All H~{\sc ii} regions in our sample have diameters less than 130 pc, but we do not find a slope of zero.  
We speculate why we do not find a zero slope as either due to the fact that we have not detected several 
large ($\sim$45 pc) H~{\sc ii} regions or that 
the resolution differences between our and past images are significant. 
The first possibility can be ruled out because our HST H$\alpha$ images have 
exceptionally high signal-to-noise and high spatial resolution.  
Such large systems would be detected easily.  
Thus, the differences in spatial resolution of HST make this assertion untrue that the power 
law slope should be flat for diameters less than 130 pc. 

Comparing our number of H~{\sc ii} regions with those from 
\citet{youngblood99}, we have 30 times more H~{\sc ii} regions.  
Our largest H~{\sc ii} region diameter is 3 times smaller (45 pc versus 159 pc) than that observed by \citet{youngblood99}, demonstrating the superior resolution of the HST images.

\subsection{H~{\sc ii} Region Luminosity Function}

Our H~{\sc ii} region luminosity function slope, -1.00, found in this study is similar to those from 
previous studies of irregular galaxies \citep[-0.60 to -2.67][]{youngblood99}. 
Even with the resolution of the HST H$\alpha$ image, we get a similar slope.  
The HST image allows us to resolve H~{\sc ii} region complexes that are 
listed in past ground-based observations \citep[e.g.,][]{waller91} into several smaller H~{\sc ii} regions.  
These large H~{\sc ii} region complexes were used by \citet{youngblood99}.  Their slope is steeper and covers 
a smaller luminosity range.  Their slope fits our luminosity data over their luminosity range of 38.0--39.0 dex.  
We derive a similar slope of -1.22 over a larger luminosity range (37.22--38.91 dex) if we only fit our histogram 
bins before they begin to turnover.  
Thus, the slope of \citet{youngblood99} is accurate to luminosities down to logarithmic H$\alpha$ luminosities of 37.22 dex.  

We can test the statement of \citet{oey98} 
that a power law slope used on luminosities below L$_{\rm sat}$ (10$^{38.5}$ ergs  s$^{-1}$) should be flat according to their models. We measured only five H~{\sc ii} regions above this saturation limit, and we do not measure a flat slope for our luminosity function.  Therefore, a nearly unsaturated (the number of stars in a H~{\sc ii} region does not sample the IMF well) sample of H~{\sc ii} regions produces a luminosity function with a shallow, though statistically significant, slope.  
\citet{oey98} also state that the slope of the luminosity function should break in two locations,  
once at the average luminosity of the luminosity function and 
once at the saturated luminosity value of 10$^{38.5}$ ergs s$^{-1}$.  
We fit a linear regression to the bins in Figure \ref{h2lf} 
that are below and above the average luminosity (10$^{37.16}$ for our results) as well as the saturation limit.
The slope below the average luminosity is 1.08$\pm$0.44, the slope above the average luminosity and below the saturated luminosity is 
-1.24$\pm$0.07, and the slope above the saturation limit is -1.55$\pm$0.34.  The significances of these fits are 55\%, 97\%, and 91\% respectively.  
The slope below the average luminosity is different than the slope above the average luminosity but not significantly.  
The other two fits are statistically significant, but the slopes are within each other's 3$\sigma$ confidence intervals.  
Since \citet{oey98} discuss the slopes by themselves, we do find that the slope above and below the average luminosity are statistically significant.  We
do not find that the slope for luminosities 
greater than the average luminosity is different that that for slopes greater than the saturation luminosity.



\subsection{Consequences of a Large Amount of Non-Photoionized Emission \label{consequences}}

Our finding implies a large amount of non-photoionized gas in a small galaxy, 
and leads us to suspect previous  derivations of reddening and metallicities based on ground-based data.  
Possibly the metallicity of this galaxy is in fact larger than previously thought.  
However, [O~{\sc iii}]/H$\beta$ and [S~{\sc ii}]/H$\alpha$ alone cannot give accurate metallicity measurements 
because the emission-line ratios are degenerate in metallicity and ionization parameter \citep{kewley01}.
Reddening could be lower than previously thought because the H$\alpha$ emission relative to H$\beta$ is increased 
by shocks and other non-photoionizing mechanisms.  
We compare the C(H$\beta$) averages of the non-photoionized and 
photoionized gas using the 3$\times$3 5$\sigma$ maps.  
The average for the non-photoionized gas is 0.76 and that of the photoionized 
points is 0.67.  
We test the significance in the differences in the averages using the F$^*$-test for means.  We calculate a p-value of 0, implying a significant difference.  
When the average C(H$\beta$) is taken for the entire galaxy, the value is 0.69.  
Thus, the presence of non-photoionized gas increases the derived C(H$\beta$) by only 0.02 (or A$_V$ of 0.04 mag) over the pure photoionized average.  Because reddening is not affected significantly by a large percentage of non-photoionized emission, we hypothesize that the metallicity will not be altered by more than 0.2 dex.

To determine if the large non-photoionized emission has any noticeable effect on the luminosity function and size distribution, 
we used HIIphot on the HST H$\alpha$ images of NGC 4214 using a similar procedure to that described in \S \ref{computelf} and \S \ref{computesd}.  
The main difference between NGC 4214 and NGC 1569  
is the amount of non-photoionized emission.  
No other differences are significantly different between these two galaxies, such as metallicity, star formation rate, etc.
These H$\alpha$ images are the ones used by \citet{calzetti04}.  
We find a statistically significant luminosity function slope of -1.30$\pm$0.07 over a 
range of luminosities from 36.90 dex to 38.64 dex (sample size of 438 objects).  
The slope is steeper than the one found for NGC 1569 over a luminosity range of 36.81 dex to 38.91 dex, -1.00$\pm$0.08.  
Neither slope is found in the 3$\sigma$ confidence interval of the other, but their confidence intervals do overlap.  
However, our other slope for the range from 37.22 dex to 38.91 dex, 
-1.22$\pm$0.06, is well within the 3$\sigma$ confidence intervals of the other. 
We find a statistically significant size distribution slope of -2.37$\pm$0.2 over a 
range of diameters from 7.28 pc to 124 pc.  This slope is 
shallower than the -3.39$\pm$1.94 slope from \citet{oey03}, but our slope for NGC 4214 is 
found well inside their confidence intervals (but not vice versa).  The 
slope is also found outside the confidence intervals of our slope for NGC 1569 and vice versa.  The maximum diameters are also a factor of 3 different, 45 pc for NGC 1569 and 124 pc for NGC 4214.  The difference in size range is not caused by a difference in distances (only a factor of 1.3).  
In this comparison of slopes and ranges to the non-photoionized emission, we can speculate that the difference in 
mechanical luminosity, the only significantly different property between these two galaxies, could cause an overpressurization in the ISM or 
strip the ISM around the H~{\sc ii} regions in NGC 1569 compared to NGC 4214.  
Either scenario could shrink the detected H~{\sc ii} region diameters in NGC 1569 compared to NGC 4214.  
The lack of difference in the luminosity function slopes suggests that the 
scenario of overpressurization in the ISM is the more likely choice.  
If the material around the H~{\sc ii} regions was being stripped, then we would see a distinct difference in
the luminosity range and luminosity function slopes 
in NGC 1569 (smaller) compared to NGC 4214 (larger).
This overpressurization scenario can be checked by determining the density of these H~{\sc ii} regions.  
The average density of the H~{\sc ii} regions in NGC 1569 should be larger than the average density of NGC 4214.

\section{SUMMARY \label{summ}}

We have taken the {\it WFPC2} narrowband imagery of NGC 1569 and generated H$\alpha$/H$\beta$, [O~{\sc iii}]/H$\beta$, and [S~{\sc ii}]/H$\alpha$ ratio maps.  We have determined the areas of non-photoionized and photoionized emission on a 3 pc by 3 pc scale.  
The non-photoionized H$\alpha$ luminosity of NGC 1569 is 10\%$\pm$3\% of the total H$\alpha$ luminosity.
This value is approximately 2.5 to 3 times larger than percentages found in similar starburst galaxies \citep{calzetti04}. However, we did point out that our method 
of determining the percentages are different from these past results.  If we use 
a method more similar to theirs, we find that our percentage is 1.5--2 times smaller
than these previous results.  Thus, we can conclude the non-photoionized percentage
is different for this object compared to others.   

One-half to two-thirds of the non-photoionized emission originates in high surface brightness areas.  We can explain the amount of non-photoionized H$\alpha$ luminosity using an instantaneous burst which uses a mass at least equal to the masses of the largest three star clusters in NGC 1569. Or we can explain the amount of non-photoionized H$\alpha$ luminosity using a continuous star formation scenario with a SFR of 0.15 M$_{\odot}$ yr$^{-1}$ for 5 Myr and then the SFR from  \citet{greggio98} for at least 10-20 Myr.  We conclude that the main non-photoionizing mechanism is shocks mainly produced by supernovae or massive stellar winds.  
Individual sources responsible for these shocks include supernova remnants, x-ray binaries, and star clusters.  The most prominent non-photoionized area is the arc/shell surrounding the brightest H~{\sc ii} region complex.
This site of vigorous star formation will likely continue to produce stars for the next several Myr.  

We have also used the H$\alpha$ WFPC2 image to generate H~{\sc ii} region size distribution and luminosity function.  
We find that a power law best describes the size distribution.  The slope derived for the H~{\sc ii} region size distribution ($\alpha = -3.02\pm0.27$) is flatter than slopes for other irregular galaxies, but fits well within the confidence intervals of the past results and vice versa.  
Our luminosity function slope agrees with the result of \citet{youngblood99} over a logarithmic H$\alpha$ luminosity range of 37.22--38.91 dex. 
We also derive a flatter slope, -1.00$\pm$0.08, that is within the norm for irregular galaxies and that is over the logarithmic luminosity range of 36.81--38.91 dex.  

The mechanical luminosity, which drives the non-photoionized emission, does not adversely affect the estimates of reddening or metallicity from  ground-based spectroscopy by more than 3\%.  
However, the slope of the size distribution as well as the range of sizes for NGC 1569 is different than that for NGC 4214.  This finding coupled with the similarity in the luminosity function slopes and ranges for these galaxies suggests that overpressurization of the ISM from non-photoionized emission
could cause these problems.  Future density measurements of these H~{\sc ii} regions could show explicitly whether this speculation is true.  
Within 4 pc of the 10-20 Myr old superstar clusters A1, A2, 
and B, no bright
HII regions exist to a luminosity limit of 2.95$\times10^{36}$ erg s$^{-1}$,
suggesting that the winds and shocks have effectively terminated star
formation in this small cavity. 
In the three annular regions around 
the SSCs, both the HII region luminosity function and HII region size
distribution are consistent with respect to one another and the galaxy as a whole.  
The HII region surface densities within the annuli remain constant as the annuli are moved
away from the superstar clusters.  
These results indicate that these feedback effects in NGC~1569 are confined to the
immediate vicinity of the most recent massive star formation event on
scales of $\sim1$ pc.

With these results, we add further evidence that a single burst event $\sim$10 Myr ago created 
the super star clusters A1, A2, and B as well as 
the other star clusters responsible for the 
dominant event producing the current non-photoionized emission and properties of H~{\sc ii} regions.

\acknowledgements

We thnak the referee for her/his thoughtful insights into making this paper better.  
We thank the following people for providing us with the multiwavelength data found in this paper: Crystal Martin \& Chris Taylor.  
We thank Lisa Kewley for providing the model results from her paper and for her helpful comments on this paper.  
B.B.\ and H.A.K.\ are supported by NASA grant NRA-00-01-LTSA-052.

\clearpage

\begin{figure}
\epsscale{1.0}
\figcaption{H$\alpha$ Images of NGC 1569.  
Both images have been binned in 3$\times$3 pixel bins and 
are displayed together on the same scale to show the locations of the 
non-photoionized points (white dots in the 
lower image).    
The upper image displays only the H$\alpha$ image of NGC 1569.
The lower H$\alpha$ image depicts the locations of the non-photoionized 
gas, exceeding the 5$\sigma$ threshold, as white dots.  Super Star Clusters
A1, A2, and B are labeled to the right of the objects.  The grayscale bars 
indicate the levels of the image and are in the units of 10$^{14}$ ergs 
s$^{-1}$ cm$^{-2}$.  
\label{hanonpi} \label{compass}}
\end{figure}

\begin{figure}
\epsscale{1.0}
\figcaption{[O~{\sc iii}]/H$\beta$ ratio map of NGC 1569.  
This ratio map has the same labeling and notation as Figure \ref{compass}.  
Each point represents the 5$\sigma$ 3$\times$3 
[O~{\sc iii}]/H$\beta$ binned results used in this study.  
The grayscale bar to the right indicates what numerical values that 
the various shades of grey represent.  
\label{o3hb}}
\end{figure}

\begin{figure}
\epsscale{1.0}
\figcaption{[S~{\sc ii}]/H$\alpha$ ratio map of NGC 1569.  
This ratio map has the same labeling and notation as Figure \ref{compass}.  
Each point represents the 5$\sigma$ 3$\times$3 binned 
results used in this study. 
The grayscale bar to the right indicates what numerical values that the various shades of grey represent.
\label{s2ha}}
\end{figure}

\begin{figure}
\epsscale{1.0}
\figcaption{Plot of  log([O~{\sc iii}]/H$\beta$) vs. log([S~{\sc ii}]/H$\alpha$) for NGC 1569.  Each point represents a 3$\times$3 bin of pixels 
(3.2 pc on a side) with a H$\beta$ value 5$\sigma$ above the 
mean zeroed background of the H$\beta$ image.  
The solid line represents the maximum starburst line from 
\citet{kewley01}.  
The position and distribution of the points is similar to the ratios 
of NGC 4214 from \citet{calzetti04}.  
However, the percentage of non-photoionized H$\alpha$ emission is greater 
than the percentage of NGC 4214, by a factor of 3.  
\label{lo3hbvls2ha}}
\end{figure}


\begin{figure}
\epsscale{1.0}
\figcaption{Wolf-Rayet star locations superimposed 
on the NGC 1569 H$\alpha$ image.  
This image has been binned in 3$\times$3 pixel bins.  
Superimposed as white squares are the $>$5$\sigma$ locations of 
the non-photoionized gas within NGC 1569.
Superimposed in gray circles with black outlines are the locations of 
Wolf-Rayet stars or star clusters with the presence of Wolf-Rayet stars.  
Each circle is labeled with its designation 
from \citet{buckalew00}.  
Orientation of this figure is the same as that
of Figure \ref{compass}. 
Note the arc directly below C5, 
indicating that the wind from this cluster is 
possibly driving into the nearby ISM and creating a localized region of non-photoionized H$\alpha$ emission.  
\label{n1569wr}}
\end{figure}

\begin{figure}
\epsscale{1.0}
\figcaption{F555W images of NGC 1569 taken from the archival data of 
GO-6111 (left) and GO-6423 (right).  
Orientation of these images are given by the compasses in the 
respective images.  
Superimposed in white are the 5$\sigma$ 3$\times$3 locations of the 
non-photoionized gas within NGC 1569.  
The numbers are the star cluster designations from \citet{hunter00} and are discussed in \S \ref{clusters}.
\label{n1569cluster}}
\end{figure}

\begin{figure}
\epsscale{1.0}
\figcaption{Supernova remnants labeled on the NGC 1569 H$\alpha$ image.  
Superimposed in white are the 5$\sigma$ locations of 
the non-photoionized gas within NGC 1569.
Superimposed as gray circles with black outlines are the locations of radio 
supernovae and supernovae remnants from \citet{greve02}.  
Each circle is labeled with its designation 
from \citet{greve02}.  
Note that most of these point sources have some nearby or 
superimposed non-photoionized gas 
(see \S \ref{snr} for more discussion).
\label{n1569snr}}
\end{figure}

\begin{figure}
\epsscale{1.0}
\figcaption{X-ray point sources labeled on the NGC 1569 H$\alpha$ image.  
Superimposed in white are the 5$\sigma$ locations of the 
non-photoionized gas within NGC 1569.
Superimposed as gray circles with black outlines are the locations of X-ray 
point sources from \citet{martin02}.  
Each circle is labeled with its designation 
from \citet{martin02}.  
Note that most of these point sources have some nearby 
or superimposed non-photoionized gas (see \S \ref{x} for more discussion).
\label{n1569x}}
\end{figure}

\begin{figure}
\epsscale{1.0}
\figcaption{Broadband X-ray image from \citet{martin02}.  
Superimposed in white are the 3$\times$3 5$\sigma$ locations of the 
non-photoionized gas within NGC 1569.  
North is up, and east is to the left.  
We have circled in gray the only obvious coincidence 
of extended X-ray emission and non-photoionized gas.  
This region is the brightest H~{\sc ii} region complex in 
the galaxy and the location of the strong thermal radio source 
M-1 of \citet{greve02}.  
No other associations are apparent.
\label{n1569chandra}}
\end{figure}


\begin{figure}
\epsscale{1.0}
\figcaption{CO map of NGC 1569 from \citet{taylor99}.  
Superimposed in black are the 3$\times$3 5$\sigma$ locations of the 
non-photoionized gas within NGC 1569.  
We have labeled the CO detections as designated in \citet{taylor99}.  
North is up, and east is to the left.  
The only CO emission objects near non-photoionized points are 3 
and 4 \citep[designations from ][]{taylor99}.  
The position of object 3 is the same as the extended X-ray emission 
circumscribed by non-photoionized points and as M-1, 
the large thermal radio source from \citet{greve02}.  
\label{n1569co}}
\end{figure}


\clearpage
\begin{figure}
\epsscale{1.0}
\figcaption{Histogram of H II region diameters 
with the fitted size distribution overplotted.  
The solid line is the fit as described by the equation in Table \ref{tab3}.  
The points are the bins as derived using the equation of \citet{scott79}.  
The x-axis uncertainties represent the size of the individual bins.  
The y-axis uncertainties are 1$\sigma$ in length.  
\label{h2sd}}
\end{figure}

\begin{figure}
\epsscale{1.0}
\figcaption{H$\alpha$ luminosity histogram with the fitted luminosity 
function overplotted.
The solid line is the fit of all luminosity bins and is 
described by the equation in Table \ref{tab3}.  
The long dashed line is the fit of the luminosity bins covering
the range from 37 to 38.91 and is described in Table \ref{tab3}.  
The points are the bins as derived using the equation of \citet{scott79}.  
The x-axis uncertainties represent the size of the individual bins.  
The y-axis uncertainties are 1$\sigma$ in length.  
The dashed line in the bottom right corner is the slope as derived by 
\citet{youngblood99}.  
This line is significant only between 38 dex and 39.9 
dex and fits our data points over this range too. 
The slope of the our long dashed line and that of \citet{youngblood99}
are similar.  
The diamonds above the four leftmost points are the incompleteness
corrected values.  Note that they do not change the shape of the 
histogram much.  
\label{h2lf}}
\end{figure}

\begin{figure}
\epsscale{1.0}
\figcaption{H$\alpha$ image with the positions of the H~{\sc ii} regions in our sample superimposed as white squares.  These H~{\sc ii} regions were 
chosen by HIIphot to have a signal-to-noise of 100 or better.  Roughly 1000 H~{\sc ii} regions were detected at this signal-to-noise threshold.  Three annuli are drawn on this image to represent where our subsamples are located.  Each subsample has the same number of H~{\sc ii} regions.  These subsamples have the same luminosity function and size distribution slopes with respect to one another, indicating no change by proximity to the super star clusters.  The surface density of H~{\sc ii} regions is 1.33 times higher in the first annuli (the one nearest the super star clusters) than that from the next annuli out.  
\label{f13}}
\end{figure}

\clearpage
\begin{deluxetable}{lccccc}
\tablewidth{0pt}
\tablenum{1}
\tablecaption{Observational Parameters of HST/WFPC2 New \& Archival Data \label{tab1}}
\tablehead{
\colhead{Filter} &   \colhead{Band/}       & \colhead{PI}      &
\colhead{GO-Program} &
\colhead{Date}          & \colhead{Exposure Time} \\
\colhead{}   & \colhead{Emission Line}      & \colhead{}      &
\colhead{Number} &
\colhead{}          & \colhead{(sec)}}
\startdata
F487N & H$\beta$      & Shopbell & 8133 & 23 Sept.\ 1999 & 3200 \\
F502N & [O~{\sc iii}] & Shopbell & 8133 & 23 Sept.\ 1999 & 1500 \\
F547M & Str\"{o}mgren $y$ & Shopbell & 8133 & 23 Sept.\ 1999 & 60 \\
F656N & H$\alpha$         & Shopbell & 8133 & 23 Sept.\ 1999 & 1600 \\
F671N & [S~{\sc ii}]      & Shopbell & 8133 & 23 Sept.\ 1999 & 3000 \\
\enddata
\end{deluxetable}

\clearpage
\begin{deluxetable}{lcc}
\tablewidth{0pt}
\tablenum{2}
\tablecaption{Measured and Derived Quantities of NGC 1569 (3$\times$3 5$\sigma$ results) \label{tab2}}
\tablehead{
\colhead{Property} & \colhead{Units}  & \colhead{Value}}
\startdata
F$_{\rm H\alpha}$ & ergs s$^{-1}$ cm$^{-2}$ & 3.4$\times10^{-11}$ \\
L$_{\rm H\alpha}$ & ergs s$^{-1}$           & 1.9$\times10^{40}$  \\
SFR$_{\rm H\alpha}$ & M$_{\odot}$ yr$^{-1}$ & 0.15                  \\
Q$_0$(H$\alpha$)   & photons s$^{-1}$      & 1.4$\times10^{52}$  \\
F$_{\rm H\alpha,non-pi}$ & ergs s$^{-1}$ cm$^{-2}$ & 5.0$\times10^{-12}$ \\
L$_{\rm H\alpha,non-pi}$ & ergs s$^{-1}$           & 2.9$\times10^{39}$  \\
L$_{\rm H\alpha,non-pi}$/L$_{\rm H\alpha,total}$ & & 0.15  \\
A$_{\rm non-pi}$/A$_{\rm total}$ &                & 0.16 \\
L$_{\rm mechanical}$            & ergs s$^{-1}$  & 0.56-1.1$\times10^{41}$ \\
L$_{\rm H\alpha,mechanical}$/L$_{\rm H\alpha,total}$ & & 0.1-0.2 \\
\enddata
\tablecomments{``non-pi" indicates non-photoionized results.}
\end{deluxetable}

\clearpage
\begin{deluxetable}{lcc}
\tablewidth{0pt}
\tablenum{3}
\tablecaption{H~{\sc ii} Size Distribution and Luminosity Function Equations \label{tab3} \label{tab4}}
\tablehead{
\colhead{Property} & \colhead{Units}  & \colhead{Value}}
\startdata
Diameter Lower \& Upper Limits & pc & 6.3--44.8 \\
Size Distribution Equation & & log(N)$=$-3.02$\pm$0.27$\times$log(D[pc])+5.16$\pm$0.27 \\
Adjusted R$^2$ & & -0.9 \\
Luminosity Lower \& Upper Limits & ergs s$^{-1}$ & 10$^{36.81}$--10$^{38.91}$ \\
Luminosity Function Equation & & log(N)$=$-1$\pm$0.08$\times$log(L[ergs s$^{-1}$])+39.25$\pm$2.14 \\
Adjusted R$^2$ & & -0.92 \\
Luminosity Lower \& Upper Limits & ergs s$^{-1}$ & 10$^{37.22}$--10$^{38.91}$ \\
Luminosity Function Equation & & log(N)$=$-1.22$\pm$0.06$\times$log(L[ergs s$^{-1}$])+47.5$\pm$2.14 \\
Adjusted R$^2$ & & -0.97 \\
\enddata
\tablecomments{1$\sigma$ uncertainties come after the $\pm$.  
Adjusted R$^2$ values are a measure of how well the linear regression fits 
the data given their uncertainties.  
The higher the number (maxing at 1), the better that the fit is. 
Significant regressions usually have values between 0.9 and 1.  
A negative value indicates an inverse correlation.} 
\end{deluxetable}

\end{document}